# Software Security Rules: SDLC Perspective

C. Banerjee, S. K. Pandey
Department of Information Technology
Board of Studies, The Institute of Chartered Accountants of India, Noida- 201301, INDIA

*Abstract*---Software has become an integral part of everyday life. Everyday, millions of people perform transaction through internet, ATM, mobile phone, they send email & e-greetings, and use word processing and spreadsheet for various purpose. People use software bearing in mind that it is reliable and can be trust upon and the operation they perform is secured. Now, if these software have exploitable security hole then how can they be safe for use. Security brings value to software in terms of people's trust. The value provided by secure software is of vital importance because many critical functions are entirely dependent on the software. That is why security is a serious topic which should be given proper attention during the entire SDLC, *'right from the beginning'*. For the proper implementation of security in the software, twenty one security rules are proposed in this paper along with validation results. It is found that by applying these rules as per given implementation mechanism, most of the vulnerabilities are eliminated in the software and a more secure software can be built.

*Keywords-Security rules, Security rules in SDLC, Software Security*

## I. INTRODUCTION

The issue related to computer security surfaced for the first time in 1970s with report of earliest known intrusion in 1977, first spam email in 1978, earliest large-scale identity theft in June 1984 and attack of first known computer virus reported in 1987. In hackers named 414s attacked 60 computer systems. During 1980s and 1990s, many international banks were targeted by crackers and hackers. In 1995, U.S. Department of Defense computers were attacked roughly 250,000 times. In 1996, hackers alter websites of the U.S. Department of Justice in August, CIA in October, and U.S. Air Force in December. In 2001, Microsoft becomes victim of Denial of Service attacks. In May 2006, a Turkish hacker successfully hacked 21,549 websites. In March 2008, around 20 Chinese hackers claim of gaining access to the world's most sensitive sites, including Pentagon. In April 2009, Conficker, a worm infiltrated billions of PCs worldwide including many government-level top-security computer networks [1] [2].

While trying to identify and analyze the reason behind the cause of security breach, we generally put blame entirely on virus attack, denial of service, spam mail etc. If we introspect in true sense, we see that our thinking becomes so partial that while analyzing the facts we intend to forgo a very important and real fact which is one of the most important factors in software security breach, and, that is, bad software which is actually behind every security problem and malicious attack [3]. Besides identifying and targeting those individual security threats and providing solution for those attacks, if we also put focus on the security aspect of software, we surely can build a more robust and reliable system in totality.

Security loop holes in software can also endanger intellectual property and business operations and services. It is estimated that 70 percent of reported security incidents result from exploits against defects in the design or code of software [4] [5]. It is a pre assumption that security features implemented in software delays the project as it adds time and increase the cost. Due to this, many designers tends to ignore or given little importance to the security aspect of the project. However, implementing security in software by complying with regulatory standards gives long term benefits in terms of litigation avoidance, protection against loss of sensitive information, and protection against loss of reputation. It also provides assurance that the data in a system has a reasonably expectation of protection and privacy [6]. It also ensures reliability, integrity, and safety for the system using secured software.

Implementing security in software from the very stages of its development makes the system as vulnerable and fault free as possible. It further enforces limits on the damages occurring consequently due to various failures caused by attack triggered fault. It also provides mechanism for quick recovery by the system from the damages caused by failure. It ensures that the system continues to operate under most adverse condition created due to the various attacks on the system. In doing so, the system provide a mechanism of resistance against the attacker who tries to exploit the weakness in the software. It also provides a tolerance level of such failures resulting from such exploits [7].

In a 2005 report, approximately 163000 consumer records were stolen leading to the case of identity thefts with a US $10 million settlement fine. In 2006, hackers accessed the account and personal information of nearly 19,000 AT&T credit card holders [8]. Estimated revenue losses due to piracy in Asia-Pacific region during 2006 were reported to be US$11.6 billion [9]. In 2007, information of around 100 million credit and debit card accounts were stolen in U.S., resulting in recovery cost estimated to be about US $16 million. In U.K., loss of personal information of around 25 million people with an estimated recovery cost of about US $500 million was reported. In 2008, 4.2 million credit & debit card numbers were stolen from a supermarket chain during the card authorization process [8].

Although researchers have done remarkable work in the field of integrating security throughout the SDLC, *'right from the beginning'*, still a major portion of work needs to be carried out in order to made software more secure and reliable. In extension to the work carried out earlier, in this paper we intend to propose twenty one security rules which if practically applied from the beginning of SDLC i.e., from requirement analysis phase will definitely contribute in secure and reliable development of software. Rest of the paper is organized as follows: in section II we discuss about 'Software Security', and in section III, 'Security Rules' are given, section IV, throws light on 'Implementation Mechanism' and Section V focuses on 'Validation and Experimental Results' with 'Conclusion and Future Work' given in section VI.

## II. SOFTWARE SECURITY

The objective of software security is to imagine about the attacker and to foresee attacker's motive and perception. Generally, software development is thought of as building software that works under normal conditions. But when the security aspect is clubbed with building software, the designer and developer focal point becomes attacker's perspective and 'how they can become a threat to the software'. After proper analysis, various mechanisms of dealing with those threats can be provided. The security can be correctly build inside software by integrated it throughout the entire software development life cycle [7].

The activity of software security can be thought of as building software which performs under intentional and unintentional malicious attack [7]. The software security should exhibit ability to defend itself and the system from the attacker's exploitation and misuse of software security loop holes [10]. Moreover, software security should have the ability to identify the deficiencies of the software development process and to identify critical threats that can make software vulnerable. Software with build-in security should reflect features like predictable execution, trustworthiness and conformance. Along with these properties, the secure software should be attack resistant, attack tolerant and attack resilient [7].

Information is a very important ingredient in software and its security can be achieved by three globally accepted properties CIA (Confidentiality, Integrity, and Availability).

- **C :** Confidentiality is prevention of unauthorized disclosure of information.
- **I :** Integrity is prevention of unauthorized modification of information.
- **A :** Availability is prevention of unauthorized withholding of information.

The main objective of confidentiality is to ensure that only authorized user can access regardless of where the information is kept and how it is accessed. Confidentiality can be maintained by mechanism like access control, password, biometrics, encryption, privacy and ethics [11]. The main objective of integrity is to safeguard the accuracy and completeness of information and processing methods from being changed intentionally, unintentionally, or accidentally. Integrity needs to be maintained for ensuring privacy, security and reliability of data and information. Integrity can be maintained by mechanisms like configuration management and auditing [11]. The main objective of availability is to ensure access of information and related assets for authorized users whenever needed. Availability can be maintained by mechanisms like data backup plan, disaster recovery plan, business continuity or resumption plan [11].

## III. SECURITY RULES

The various issues encompassing software security is a point of discussion and debate among the researchers and security practitioners. One obvious way to spread software security knowledge is to train software development staff on critical software security issues. Beyond awareness, more advanced software security training should offer coverage of security engineering, design principles and guidelines, implementing risks, design flaws, analysis techniques, and security testing. Researchers have done tremendous job in this direction but there are so many research issues that need to be addresses. On the basis of various best practices available in the literature, twenty one security rules are proposed which are discussed in this section and visually shown in Figure 1.

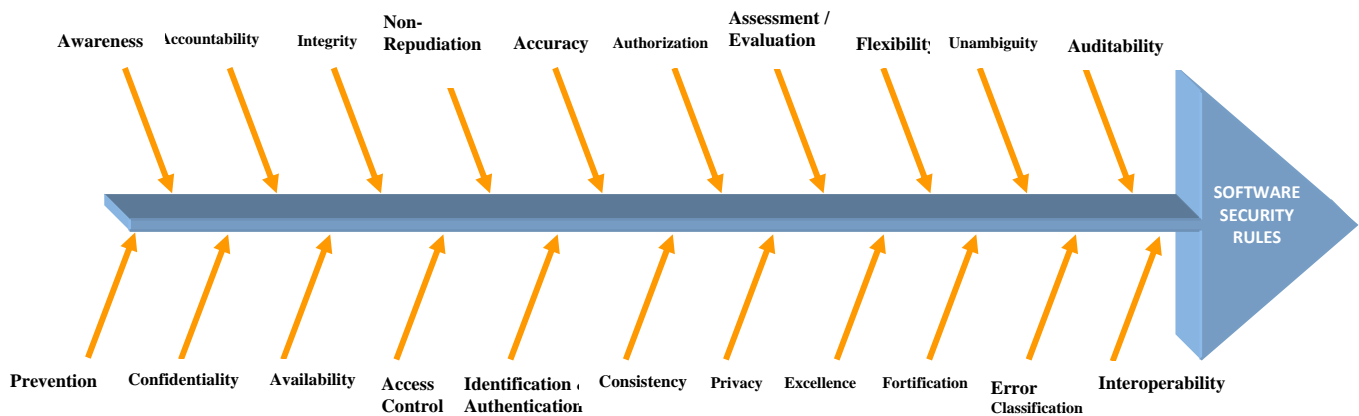

Figure 1. Rules of Software Security in SDLC

All stakeholders in software development must obey these rules in order not to introduce vulnerabilities into the system and ensure the production of secured software system. By analyzing the implementation results, it is observed that if the software engineers have these rules at the back of their minds throughout the stages of the software production, it will ensure efficient production of secure software product to a greater extent. These rules are given as follows:

**1. Rule of Awareness:** Awareness of the software security is a major point of discussion & concern among the various researchers and security practitioners [12]. *The rule suggests a constant acquisition of new information and updation of existing knowledge relating to security aspect for the software development team which includes software architecture, software developers and software testers* [13]. This can be implemented by developing an active security awareness program for training software development team on critical software security issues [12].

**2. Rule of Prevention:** As said that prevention is better than cure, in the same way, the software designer should design the software and associated security in such a way that the software when attacked internally or externally by some threat should provide some kind of safeguard and protect it from being infected. *The rule suggests that the security in software should be synchronized in such a way that it should be able to prevent any kind of threat from internal as well as external source rather then let it happen and later on cure it.* The latter option of cure is also one of the remedy but it is quite possible that by the time remedy comes into effect some more resources and application will become infectious by the infected source.

**3. Rule of Accountability:** Accountability is a key security goal which is very vital with regards to internal systems of security and reveals what a subject actually did. *The rule of accountability suggests that a log needs to be maintained for all the tasks / activities / acts performed during an operation / action with the purpose of prevention of the security policy violations and enforcement of certain liabilities for those acts* [14]. Accountability involves tracking of activities of users as well as processes and maintains their details in a log book. The main purpose of accountability is to determine the attacker or source of attack incase transaction is committed successfully [15].

**4. Rule of Confidentiality:** Security in terms of software is defined as the prevention of or protection against access to information by unauthorized persons [16]. *The rule suggests that confidentiality should be maintained by ensuring that information is not accessed by unauthorized persons* [16]. In other words, we can say that, the confidentiality in software can be maintained by keeping the contents of a transient communication or data on temporary or persistent storage secret [15]. It provides assurance that the information is shared only among authorized users or organizations [17]. The data should be handled in an adequate manner to safeguard the confidentiality of the information concerned [18].

**5. Rule of Integrity**. Software Security in respect of integrity security is the prevention of, or protection against intentional but unauthorized destruction or alteration of that information [16]. *The rule suggests that integrity should be maintained by ensuring that information is not altered by unauthorized persons in a way that is not detectable by authorized users* [16]. It provides assurance that the information is authentic and complete [17]. The integrity of data means that it can be trusted and relied upon and not that the data is 'correct' [17].

**6. Rule of Availability:** Availability is typically thought of as a performance goal, but it needs to be thought of as a security goal as the loss of availability is referred to as "denial-of-service" [15]. *The rule states that a balanced approach needs to be maintained between security and availability providing a system that is highly secure and available at all the times* [15]. It provides assurance that the systems responsible for delivering, storing and processing information are accessible when needed, by those who need them [16]. A system can ensure availability through redundancy providing alternative paths and methods in which the system is operational and functional at a given moment [16].

**7. Rule of Non-repudiation:** In general, the concept of ensuring that parties involved in a transaction can not repudiate (reject) or refute the validity of the transaction. *The rule states that the objective of non-repudiation is to ensure undeniability of a transaction by any of the parties involved where a trusted third party can play an important role* [15]. Non-repudiation protocols can be used as a tool of security to prove that the transaction actually took place and that the two parties actually interacted with each other where both the parties can not deny this fact in presence of a valid set of evidences [15].

**8. Rule of Access Control:** Access control provides a form of authority to control access to areas and resources in a given domain thereby contributing to security issue in a software development process. *The rule suggests that access to resources and services should be permission based and the user if given permission should be permitted / allowed to access those resources and services and these eligible users should not be denied access to services that they legitimately expect to receive* [16]. To have a secure software, implementation of access control in totality is mandatory.

**9. Rule of Identification & Authentication:** Authentication is the act of establishing that the claims made by a user are true which includes conforming the identity and origin of the user for security purpose [19]. *The rule suggests that the process of identification and authentication must be implemented to determine who can log on to a system and their legitimate association which various users with respect to their granted access rights* [19]. A wide variety of techniques are present to provide authentication which may include use of passwords, biometric techniques, smart cards, certificates, etc [20].

**10. Rule of Accuracy:** Assurance of accuracy in security is necessary for the software system to be secured and reliable [21]. *The rule suggests that the software development team should perform the various actions, activities, methods, process & tasks correctly and accurately every time* [13]. Here, timely accuracy is also very important from strategic point of view [22]. Highest standards of technical accuracy are also a prerequisite in designing and developing secure and reliable software.

**11. Rule of Consistency:** Consistency is an essential feature of software security which the protocol designer should keep in mind during protocol designing phase of the software. *The rule suggests that the various requirements, protocols or standards or policies designed for securing the software system should be consistent in any case.* Consistence among various security policies is a demand for secure software. Consistency should be maintained at all cost among the software system, their security requirements and violation related modules.

**12. Rule of Authorization:** Authorization is the process of verifying that an authenticated subject has the authority to perform a specified operation for security reasons [19]. *The rule suggests that the process of authorization must be implemented to determine what a subject can do on the system.* By implementing the process of authorization, it can be determined whether an identity is permitted to perform specified action or not [20]. The process of authorization can only be performed after the process of authentication has successfully accomplished [19].

**13. Rule of Privacy**: Privacy can be seen as an art of being concealed / secluded / isolated from the presence or view of others. Privacy as a social and legal issue has for a long time been a matter of concern and individual's privacy in this electronic age is increasingly endangered [23]. *The rule states that privacy can ensures that individuals maintain the right to control what information is collected about them, how it is used, who has used it, who maintains it, and what purpose it is used for [16].* Privacy protection as a tool of security can be implemented by designing and enforcing sound privacy and data protection laws and technologies [23].

**14. Rule of Assessment / Evaluation:** Assessment is a characteristic which can be applied on process or processes to get a quality software product. *The rule suggests that each and every process irrespective of size should be evaluated and assessed after it has been created by the software developer* [13]. The consistency of an assessment done for process or processes ensures the reliability of a software system [24]. Assessment is also important for the software to be valid as it measures the expected / desired output with the observed output [24]. If the assessment of a process or processes is done properly it means it is consistent and valid which represents quality a subset of security. Assessment or evaluation if done considering the current security environment can help the software developer to analyze and measure the level of security implementation in their software product versus industry standards and best practices [25].

**15. Rule of Excellence:** Quality in a software means that solution provided by that software should exactly and in totality match the needs and demands of the environment and its users. *The rule suggests that security is a subset of quality and the control and variability of the security features will depend on the quality [13].* Hence in order to achieve security in totality, the quality of the software should be of highest standards.

**16. Rule of Flexibility**: Flexibility in relation to secure software development can be defined as the systems design synchronization with security in such a way that it can adapt to the external changes when it occurs. *The rule suggests that the various requirements regarding security should not be rigid and must be flexible as well as realizable [13].* Here the details of the security specifications must be realized by the software designer and developer.

**17. Rule of Fortification (Protection):** Integrity is an important ingredient of software and it should be maintained throughout the software engineering process while implementing security for strengthening the software. *The rule suggests that the various process used in security engineering process should be secured in individuality and totality [13].* Only the concerned individual should have access to the technicalities of software security and for the rest it should be kept a secret.

**18. Rule of Unambiguity**: Unambiguity in software security means that the implementation issues of security in software should be free from anonymity and easy to understand under any circumstances by its designer and developer. *The rule suggests that for easy implementation of software security, the details pertaining to it should be clear and concise [13].* All issues related to software security must be clearly understood by the software designer and developer.

**19. Rule of Error Classification:** Security vulnerabilities very often occur due to bad error handling and due to lack of proper understanding of various errors [15]. Software developers and software security practitioners should be concerned about the various errors which create problems leading to software vulnerability. *The rule suggests that errors should be categorized & classified according to a schema containing a set of security rules for better understanding of the problem which might have an impact on the security of the software* [26, 27]. It further suggests that any error when recognized should be removed as soon as possible and should not, in any case, resurface again [13].

**20. Rule of Auditability:** Auditing in security is a feature which produces a sequential record of all the activities performed in / by a system which further aids in the reconstruction and examination of the sequence of events and/or changes in an event. *The rule suggests that auditability must be implemented to judge the accountability feature of software security and aids in redesign a full proof security policy and procedures for implementing a secure* *software system* [28]. It helps the security auditor to thoroughly understand the flow of information and develop a plan for properly securing the system. It establishes the role of security auditor as that of a validator and advisor [29].

**21. Rule of Interoperability:** In today's age, most of the software that comes in the market is platform independent and provides interoperability i.e., one software can interact with many software for exchange of data and information and for other operations. In doing so it is highly likely that software which is not secure can infect other software despite of the fact that the latter software is secured software. *This rule suggests that if more than one software are interacting or communicating with each other then all the software involved in the interaction or communication must be secured.*

### IV. IMPLEMENTATION MECHANISM

For implementing security right from the requirements phase, all the personnel from requirement engineers to maintenance engineers and other stakeholders should make themselves aware about the latest software security issues, especially the critical ones. For SDLC team, this awareness should be more technical, and, for other stakeholders, the awareness should be more general, but necessary. The requirements engineer, system software designer, programmer, test engineer, implementation engineer and the maintenance engineer should carry on their respective roles keeping in mind all the twenty one security rules quoted above. Further, the implementation engineer and the maintenance engineer should make themselves more focused on auditability and interoperability rules.

If the security rules are followed properly, it will help the requirement engineers to implement the most appropriate security mechanisms like threat modeling for meeting the true underlying security requirements. The designers will be able to design more secure design architecture and the programmer will be able to develop techniques for producing more secure coding. These security rules will broaden the role of test engineers and they will be able to choose the appropriate tool and techniques for testing the software from security point of view with focus on destructive testing. Following these security rules, the implementation engineer will be able to configure and run the software more securely. The maintenance engineers will be able to make secure maintenance plan and will help him / her to adapt the software to a more secured modified environment. The implementation mechanism of our software security rules throughout the SDLC, *'right from the beginning'* is shown in Figure 2.

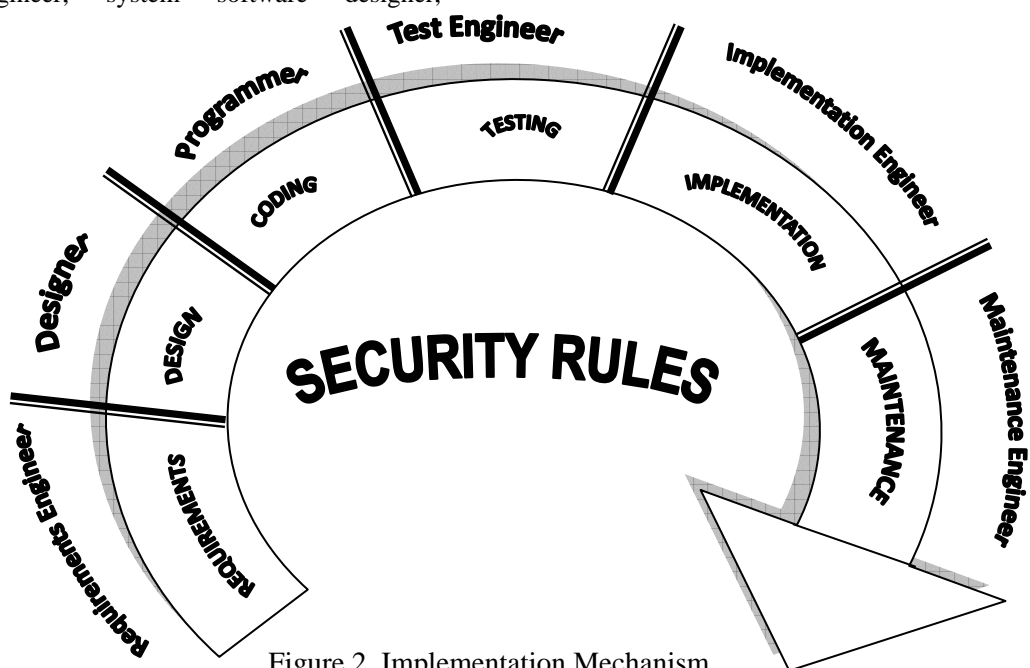

Figure 2. Implementation Mechanism

### V. VALIDATION AND EXPERIMENTAL RESULTS

These security rules were applied to a real life project from industry (on the request of the company, identity is concealed), and the final result of security assessment is calculated as per prescribed implementation mechanism. Then the level of security assurance is compared with the other project's security assurance in which these rules were not applied. The study shows that the level of risk is minimized upto 40.5%. Due to the page limit constraint, we are not providing the details of validation results in this paper; we will discuss in our next paper.

### VI. CONCLUSION AND FUTURE WORK

Secure software does not happen by accident. It is accomplished only when every designer, developer, tester and manager working on a project takes security seriously and that too during each and every phase of software development lifecycle. Security is not something that is addressed at the end of the product lifecycle nor is it a specific milestone that occurs during project execution. Security must be everywhere. It should begin at the requirement level and should be on the mind of every personnel during the entire SDLC. The paper tried to present some concrete work on software security and hence, twenty one security rules are proposed. Validation results show the applicability of these rules during the development life cycle.

Future work may include the sub division of each of the twenty one rules into their sub rules. Then set theory may be applied on those sub rules to quantify the values as well as steps. This may increase the accuracy level of these rules. These rules are validated on a project given in the validation section. Further work may be done by applying these rules on a large sample for finding the accuracy of the same. This work will help security experts to introduce security 'r*ight from the beginning*' and for building more secure software.

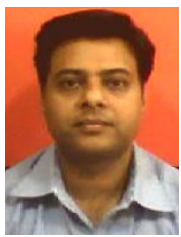
Chitreshh Banerjee is currently working as Faculty (Executive Officer) in the Department of Information Technology, Board of Studies, *The Institute of Chartered Accountants of India (Set up by an Act of Parliament) New Delhi*. Before joining the present institute, he was associated with Gyan Vihar University, Jaipur as a senior faculty. During this tenure, he was instrumental in development of Management Information System (MIS) of the university. He has an excellent academic background with a very sound academic and research experience. Under the Institute-Industry linkage programme, he delivers expert lectures on varied themes pertaining to IT. As a prolific writer in the arena of Computer Sciences and Information Technology, he penned down a number of books on Multimedia Systems, Information Technology, Software Engineering, and E-banking Security Transactions. He has contributed various research papers in the conferences of national repute. His area of interest includes multimedia systems, e-learning, e-banking, and software security.

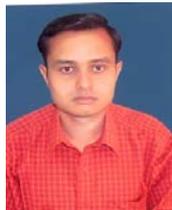
Santosh K. Pandey is presently working as a Faculty (Executive Officer) in the Department of Information Technology, Board of Studies, *The Institute of Chartered Accountants of India (Set up by an Act of Parliament) New Delhi*. Prior to this, he worked with the Department of Computer Science, *Jamia Millia Islamia (A Central University)* New Delhi. He has a rich Academics & Research experience. His research interest includes: Software Security, Requirements Engineering, Security Policies and Standards, Software Engineering, Access control and Identity Management, Vulnerability Assessment etc. Currently, he is working in the areas of Software Security and Requirements Engineering. He has published around 26 high quality research papers in various acclaimed International/ National Journals and reputed Conferences/Seminars. He has been nominated in the board of reviewers of various international/ national Journals/Conferences. His one of the research papers was adjudged as the Best Paper in the National Conference on IT- Present Practices and Challenges held at New Delhi during Aug 31- Sep 1, 2007.